\documentclass[10pt, twocolumn]{article}
\usepackage{color,soul}
\usepackage{enumitem}
\usepackage[margin=1.0in]{geometry}
\usepackage[font=footnotesize]{caption}
\usepackage{wrapfig}
\usepackage{graphicx}
\usepackage{caption}
\usepackage{subcaption}
\usepackage{stfloats}
\setlist{noitemsep}

\title{Which Sustainable Software Practices and Tools\\ Do Scientists Find Most Useful?}

\author{
    Jory Schossau \\
    \small{Michigan State University} \\
    \small{jory@msu.edu}
    \and
    Greg Wilson \\
    \small{Mozilla Science Lab} \\
    \small{greg@mozillafoundation.org}
}

\begin{document}
\maketitle

\begin{abstract}
We studied scientists who attended two-day workshops on basic software skills
to determine which tools and practices they found most useful.
Our pre- and post-workshop surveys showed increases in self-reported familiarity,
while our interviews showed that participants found learning Python more useful than learning the Unix shell,
that they found pointers to further resources very valuable,
and that background material---the ``why'' behind the skills---was also very valuable.
\end{abstract}

\section{Introduction}

Education about sustainable software practices
is key to actual adoption.
Unfortunately,
discussion about sustainable practices with the average scientist often founders
because most scientists aren't using even \emph{basic} practices \cite{b:hannay2009,b:prabhu2011}.
First encounters must therefore typically focus on concepts such as
when and where to use a script instead of a spreadsheet,
how to automate file manipulations typically performed by hand,
why to use a version control system to track work,
and how to write code for sanity and posterity \cite{b:wilson2013}.

Teaching these topics is the main aim of Software Carpentry,
a volunteer organization through which scientists (and others)
with the right knowledge and teaching skills can teach other scientists.
Its two-day workshops,
typically held at universities,
are just long enough to introduce learners to the handful of core competencies mentioned earlier.
Participants are typically excited about the subjects they learn,
but which practices are actually useful for these scientists in their daily work,
which do they actually adopt,
and how effectively Software Carpentry is teaching them?
This study's goal was to find answers to those questions.

\section{Surveys}

In the summer of 2013
the lead author designed a survey
to be administered to attendees directly before and after a Software Carpentry workshop.
This survey has let us gather information about
how well the workshop prepared attendees to handle sample real-world problems in the core competency areas,
so that we could see if what we set out to teach was being taught.
We decided early on to include only one skill-related question per core competency;
while this means we gather less (and less accurate) information per subject,
both research and experience have shown that shorter surveys are more likely to be completed.

\begin{figure}[!b]
\includegraphics[width=\linewidth]{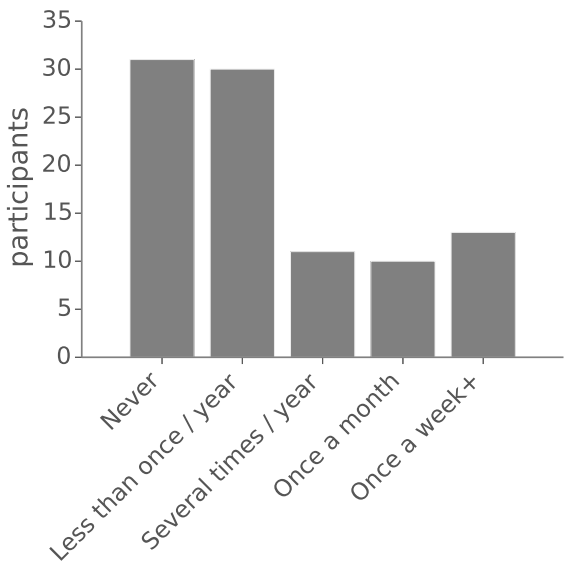}
\caption{Participants categorized by self-reported frequency of programming (n=95).\label{frequency}}

\end{figure}

The first set of questions on the survey was unique to the the pre-workshop survey.
These questions covered basic demographics such as career stage,
time spent programming,
and operating system preference,
and were used to help tailor workshops to the audience.
These questions revealed that most of our learners had little to no experience programming,
but that particular workshops would sometimes be filled with people who were more proficient.
It has proven very useful for instructors to know even this much about their learners (Figure~\ref{frequency}).



The second set of questions asked about familiarity with a topic,
and then,
if applicable,
about self-reported ability to solve a small real-world problem.
The difference between the pre-workshop and post-workshop survey responses allowed us to measure the effect of the workshop.
However,
because asking people the same question twice creates a recognition bias,
the post-workshop survey was created with similar but different questions.

\begin{figure}
\centering
\begin{subfigure}[t]{0.47\columnwidth}
    \centering
    \makebox[\textwidth][c]{\includegraphics[width=1.1\columnwidth]{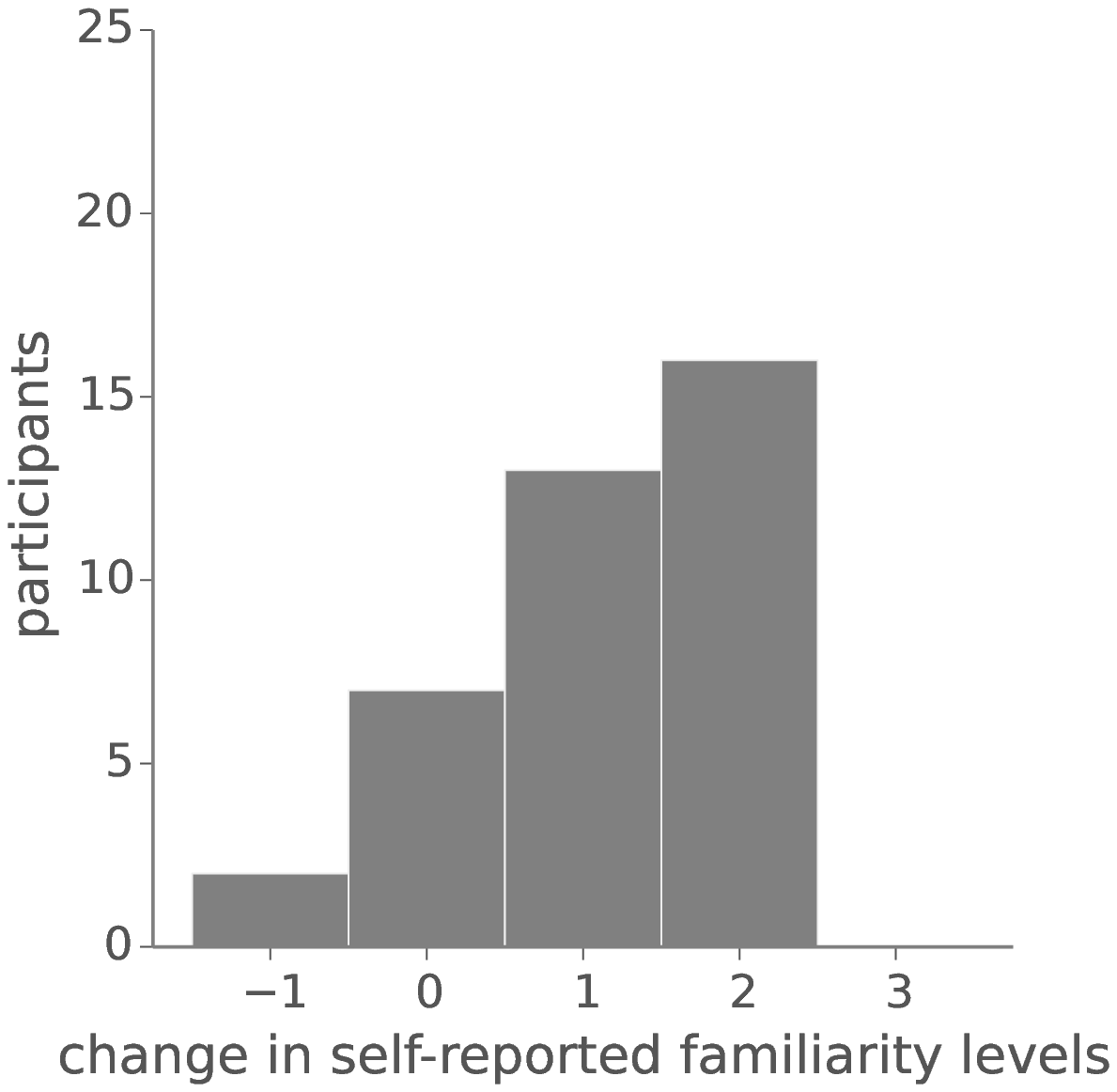}}
    \caption{Shell Familiarity: Increases in self-reported familiarity (n=43).  \label{shellFamiliarity}}

\end{subfigure}
\begin{subfigure}[t]{0.47\columnwidth}
    \centering
    \makebox[\textwidth][c]{\includegraphics[width=1.1\columnwidth]{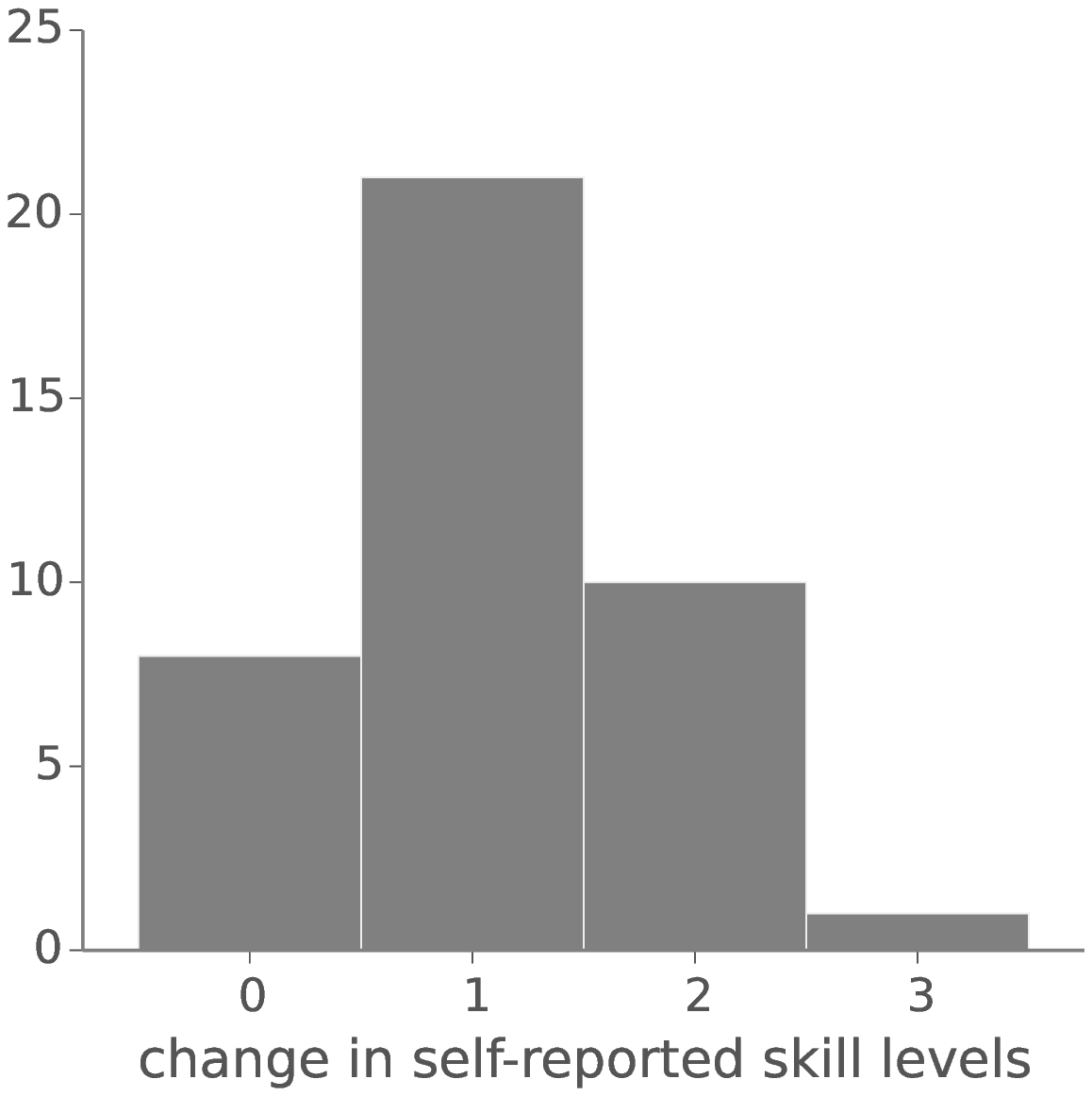}}
    \caption{Shell Skill: Increases in self-reported skill (n=40).  \label{shellSkill}}

\end{subfigure}
\caption{
    \textbf{(a)} The difference between familiarity levels before and after the workshop ($n=43$). 
    Options were ``I am not familiar with the command line,'' 
    ``I am familiar with only the term `command line','' 
    ``I am familiar with the command line but have never used it,'' 
    and ``I am familiar with the command line because I have or am using it.'' 
    The histogram represents ordinal changes. 
    The negative recordings belong to individuals
    who likely had a poor workshop experience.
    \textbf{(b)} Difference between predicted skill levels 
    before and after the workshop on similar hypothetical real-world problems ($n=40$). 
    The question posed was if the participant could perform a filename query by file contents on the command line. 
    Options were ``I could not create this list,'' 
    I would create this list using:, 
    ```Find in Files' and `copy \& paste','' 
    ``basic command line programs,'' 
    ``a pipeline of command line programs.''
    }
\label{shellFamiliarityAndSkill}
\end{figure}

\begin{figure*}[t]
\centering
\begin{subfigure}[c]{2.1in}
    \centering
        \makebox[\textwidth][c]{\includegraphics[width=2.4in]{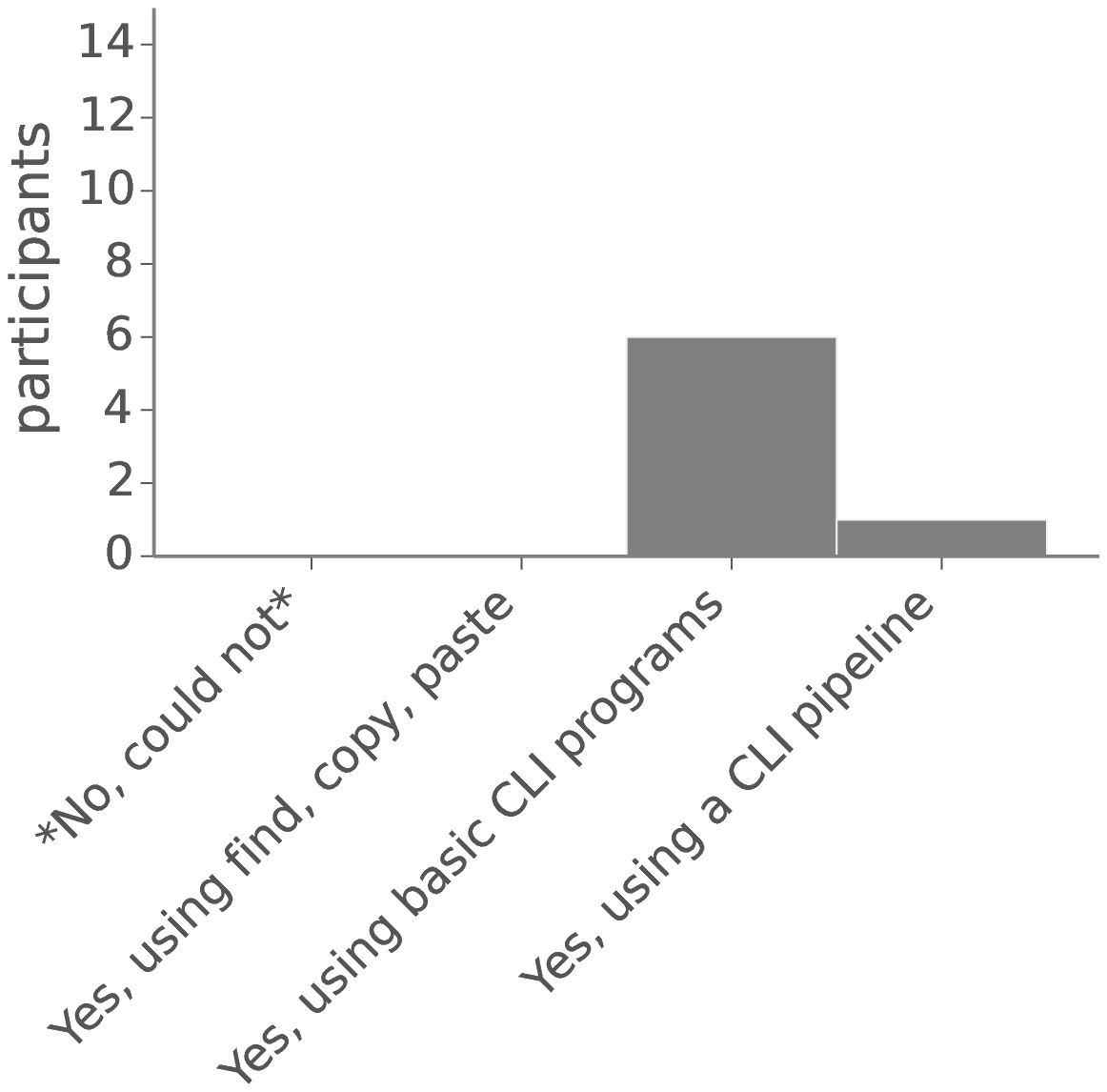}}
        \caption{
        Improvement from beginning with no ability. ($n=7$)~~~~~~~~~~~~~~~~~~~~~~~~~~
                \label{shellDetailSkill1}}

    \end{subfigure}
    \begin{subfigure}[c]{2.1in}
        \centering
        \makebox[\textwidth][c]{\includegraphics[width=2.4in]{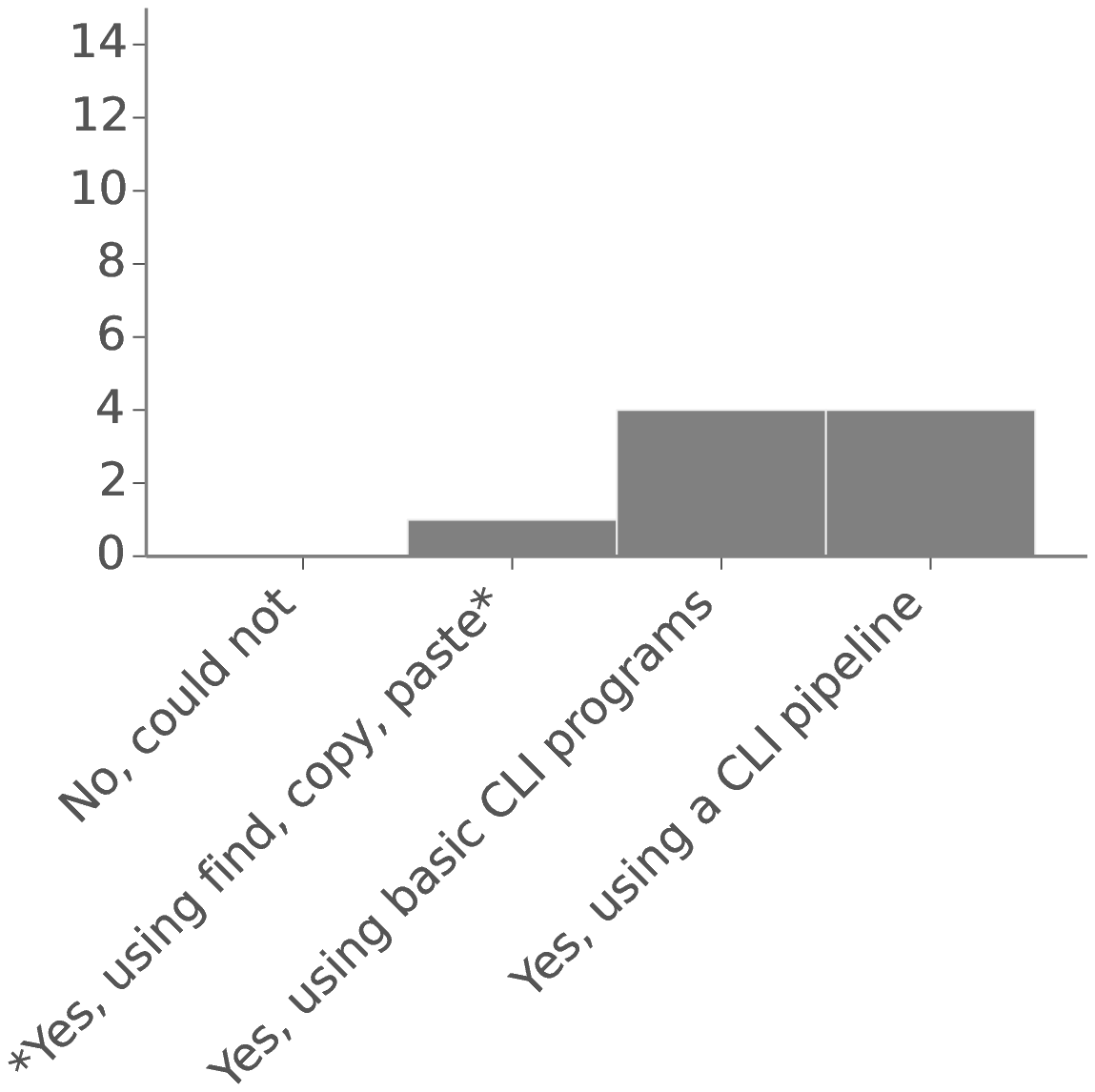}}
        \caption{
        Improvement from beginning with copy \& paste ability. ($n=9$)~~~~~~~~~~~~~~~~~~~~~~~~
            \label{shellDetailSkill2}}
    
    \end{subfigure}
    \begin{subfigure}[c]{2.1in}
        \centering
        \makebox[\textwidth][c]{\includegraphics[width=2.4in]{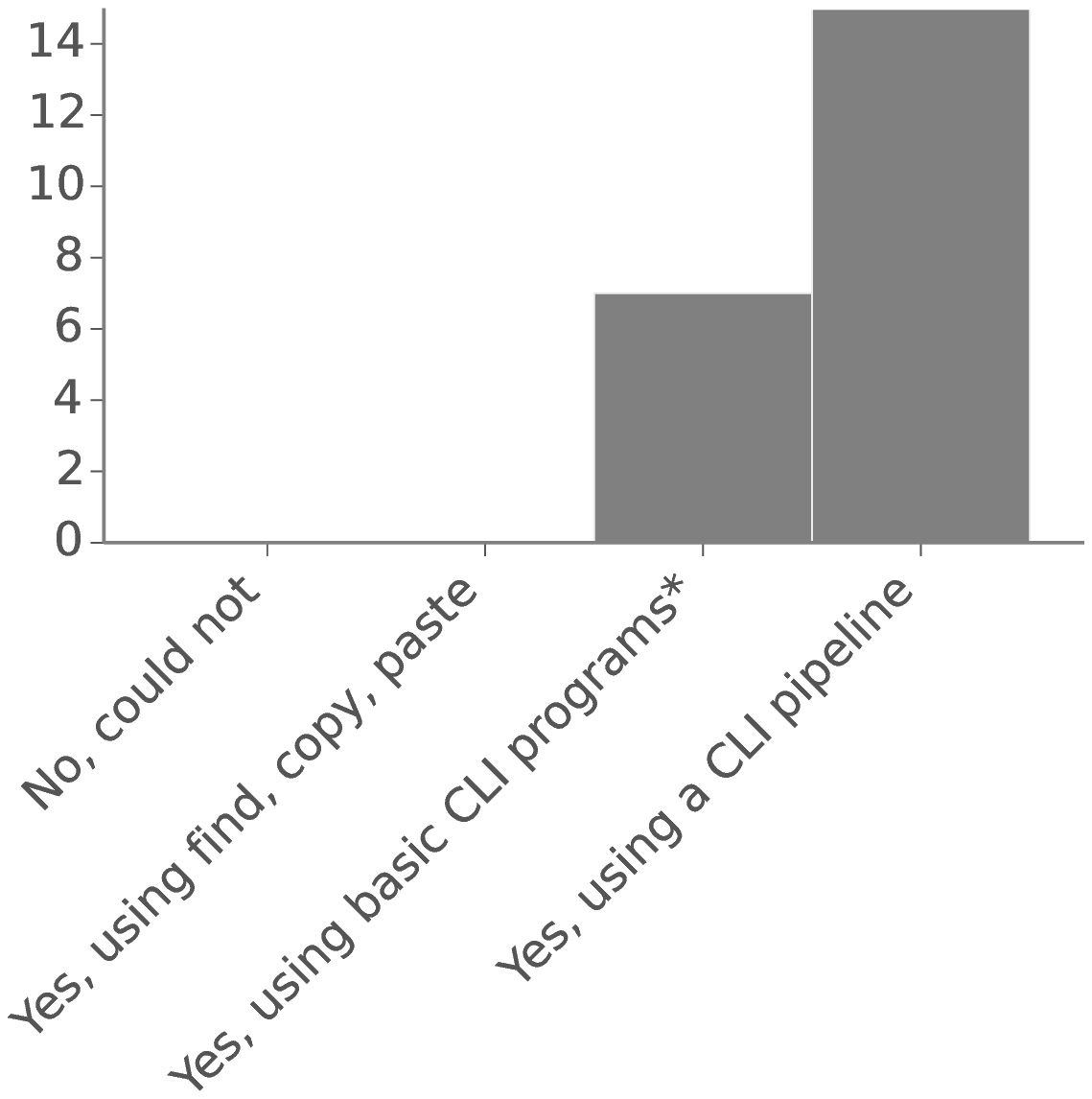}}
        \caption{
        Improvement from beginning with simple CLI commands ability($n=24$).
        \label{shellDetailSkill3}}
        
    \end{subfigure}
\caption{
  The histograms show post-workshop self-assessed ability for a real-world Unix shell problem
  given each pre-workshop self-assessed ability level represented by each figure.
  Each participant for these data answered they are familiar with the shell because they have or are using it in their work.
  This is a more detailed view of the data from Figure~\ref{shellFamiliarityAndSkill}.
\label{shellDetailSkill}}

\end{figure*}

Figure~\ref{shellFamiliarity} shows changes in participants' pre-workshop familiarity with using the Unix shell.
We expected this distribution to be highly right-skewed because
many attendees didn't have a computational background,
and workshops involved hands-on lessons with the shell.

Figure~\ref{shellSkill} shows changes in Unix shell skill level over the course of the workshop.
As we expected, most participants were able to improve their ability to use the Unix shell,
up to our maximum measure
which was the ability to use a single line of piped commands to solve the presented problem.
Overall, most participants improved one ability level,
but there were some who improved 3 ability levels,
indicating a change in self-assessment from not being able to computationally solve the problem
to self-assessing the participant could solve the problem using a single line of piped shell commands.
Figure~\ref{shellDetailSkill} shows a more detailed view of Figure~\ref{shellSkill},
separating participants by their beginning self-assessed skill level
and showing their final self-assessed skill level.
This data suggests what could be expected, 
that final achieved skill level is dependent on the beginning skill level,
but also that those who benefit most from these workshops
already have some conceptual framework of the Unix shell
and are therefore able to rapidly and effectively learn
the new information.

\begin{figure}
\centering
\includegraphics[width=\linewidth]{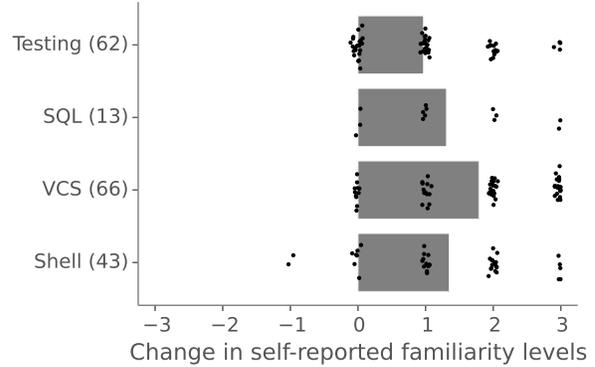}
\caption{
    Differences in self-reported familiarity for the four measured topics.
\label{AvgFamiliarityEffect}}

\end{figure}

\begin{figure}
\centering
\includegraphics[width=\linewidth]{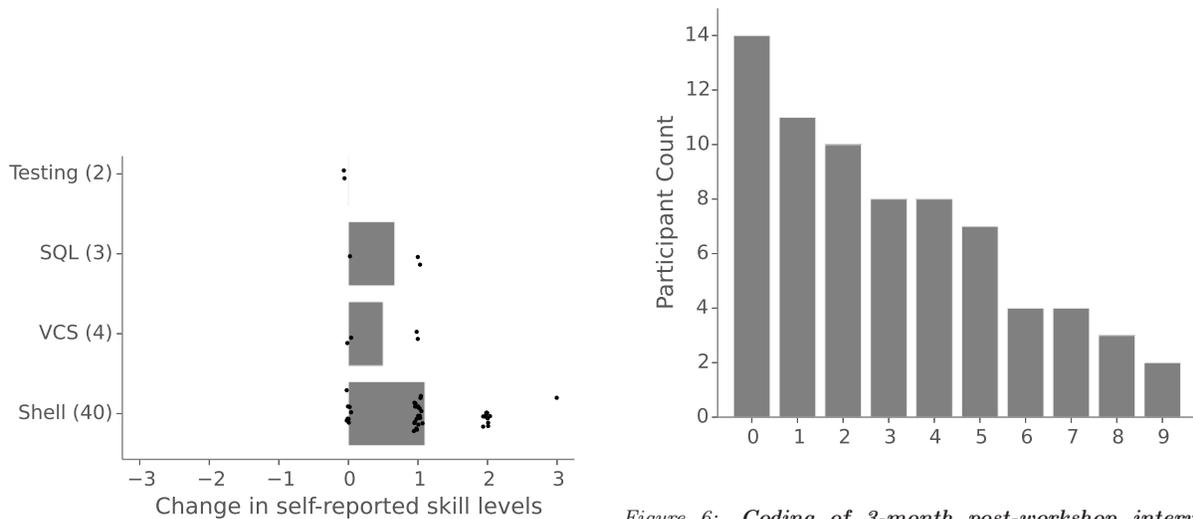}
\caption{
    Differences in self-reported skill competency for the four measured topics. Categories from which this delta was measure included: "Could not complete the task," "I could complete the task with documentation or search engine help," and "I could complete the task with little or no documentation or search engine help." Testing, SQL, and VCS allowed for a maximum of 2 skill level changes, while the Shell assessment allowed for a maximum of 3 skill level changes.
\label{AvgSkillEffect}}

\end{figure}

Figure~\ref{AvgFamiliarityEffect} summarizes the same analysis as Figure~\ref{shellFamiliarity}
for all measured topics: Unix Shell, SQL, Version Control (VCS), and Testing.
There are responses that show no increase such as for the Testing familiarity.
This is likely because Testing is sometimes left as a final topic,
leaving time to determine how in-depth the instructor may delve. 
Similarly, educational effect for skill competency
shows few data points for several topics. Figure~\ref{AvgSkillEffect}

\section{Interviews}

At the same time as we began gathering survey data,
the first author developed and began using a semi-structured protocol for interviews
in order to understand what practices scientists were finding useful six months after a workshop.
24 participants were interviewed concerning their work,
workshop experience,
how Software Carpentry helped them,
and how they might change the experience.
The transcripts of these interviews were analyzed
using the standard qualitative technique of open, axial, and selective coding,
which helped determine how best to translate qualitative responses into quantitative categorical data.

\begin{figure}
\centering
\includegraphics[width=\linewidth]{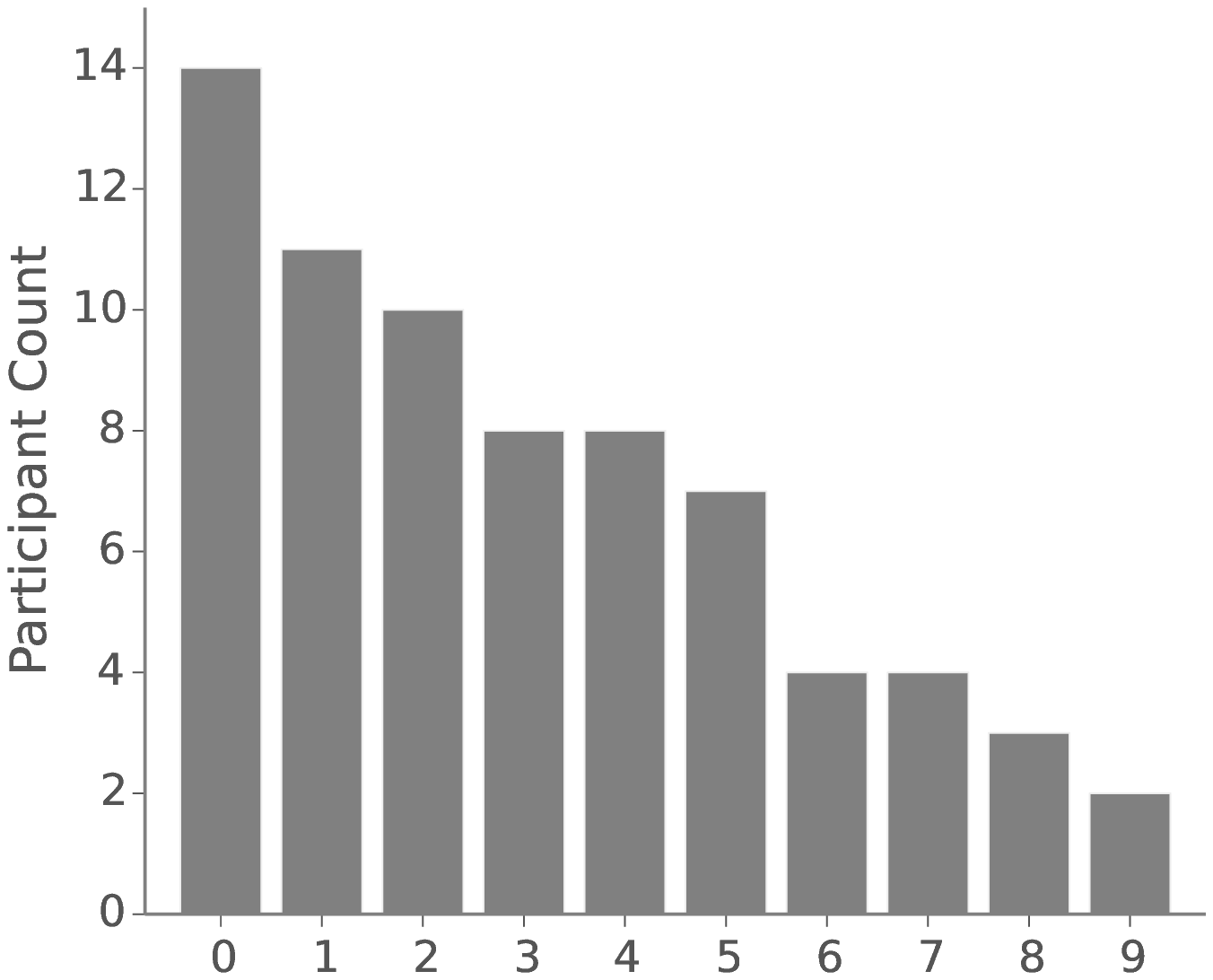}
\caption{
    Coding of 3-month post-workshop interview transcripts ($n=71$).
    Categories represent instances a participant mentioned
    directly benefiting from the given topic.
}
\label{InterviewResponses}
\end{figure}

We were initially surprised that Python was mentioned most often as the most useful topic rather than the command line.
The other core competencies followed in popularity of helpfulness,
which was more or less expected,
but some other fairly popular categories emerged during the coding process which were completely unanticipated.
These categories had been labeled ``knowing'', ``psychology'', and ``resources''.

\begin{itemize}

    \item
    \emph{Knowing} represents the satisfaction attendees have expressed that they had been told about a technology or method
    for which they had no immediate use but believed it may be useful.
    This can include simply knowing about a core competency without any planned use,
    but also includes knowing about certain software or how types of scripting work.
    
    \item
    \emph{Psychology} represents the usefulness attendees have expressed of learning about the psychology of programming,
    including empirically validated insights about how people work with computers,
    need breaks,
    or should write code.
    The second author (who taught several of the workshops from which survey participants were drawn)
    frequently cites this information while teaching a workshop.
        One subject said,
        ``Most important thing I think was the insight into the psychological basis of working.''
    
    \item
    \emph{Resources} represents attendees' mention of learning about books, websites, videos, and other sources of information
    not directly used within the workshop,
    but explicitly highlighted for their educational value.

\end{itemize}

Some participants greatly benefited from the workshop saying,
for instance,
``version control I knew nothing about and now I use regularly,''
and,
``seeing how to do things reproducibly and more efficiently was very useful,''
but others wanted more:
``too short in 2 days, so perhaps less ground and more time.''

One participant's story of benefit from Software Carpentry stood out.
Their neighborhood was being repeatedly targeted by a small group of thieves,
so they decided to use their newfound skills from Software Carpentry to build an inexpensive surveillance camera.
They used an Arduino computer with storage and camera,
and created a Python script to manage the device,
all versioned with Git.
The footage obtained from this system allowed local law enforcement to catch and sentence the thieves.
We obviously can't expect everyone to benefit quite this much,
but this example does show how broadly useful even two days' worth of training can be.

The most unexpected result from our study was a nearly unanimous opinion about what should change about the workshops:
more depth,
more exercises,
and more about working with data files and plotting.
This finding about the need for scientists to know more about how to simply work with data
falls into a long line of informal and formal feedback,
and has led to the launch of a sibling project called Data Carpentry.

\section{Future Work}

As well as surveying workshop participants,
we are now collecting data from our instructors
to find out what technologies they used,
what topics they covered,
how they think it went,
and how well they think the least skilled attendee could perform on the hypothetical problems.
We will match this data with how the participants answered to see if there was anything unusual,
such as a disconnect in perceived instructional pacing,
or to see if there are trends of topics that are difficult for participants to grasp
and difficult for the instructors to measure their comprehension.

Interviews give us a lot of insight,
but are labor-intensive and hence unscalable.
Our next goal is therefore to construct and validate better pre- and post-workshop questionnaires
so that we can track skill uptake and participant satisfaction
more systematically on a larger scale.

\subsubsection*{Acknowledgments}

Our thanks to everyone who took part in our surveys and interviews,
and to the Mozilla Science Lab and the Sloan Foundation for their support.

{\small
    \bibliographystyle{unsrt}
    \bibliography{wssspe2014-assessment}
}

\end{document}